\documentclass{www2010-submission}
\usepackage{color}
\usepackage{url}
\usepackage{graphics}
\usepackage{subfigure}
\usepackage{times}

\newcommand{\hide}[1]{}
\newcommand{\xhdr}[1]{\vspace{1.7mm}\noindent{{\bf #1.}}}
\newcommand{\xproof}[1]{ {\noindent {\it Proof.} {#1} \rule{1mm}{1mm} \vskip \belowdisplayskip} }

\newtheorem{theorem}{Theorem}

\enlargethispage{\baselineskip} 

\newcommand{\denselist}{ \itemsep -5pt\topsep-10pt\partopsep-10pt }

\begin{document}

\title{Predicting Positive and Negative Links\\ in Online Social Networks}

\numberofauthors{3}

\author{
\alignauthor Jure Leskovec\\
\affaddr{Stanford University}\\
\email{jure@cs.stanford.edu}
\alignauthor Daniel Huttenlocher\\
\affaddr{Cornell University}\\
\email{dph@cs.cornell.edu}
\alignauthor Jon Kleinberg\\
\affaddr{Cornell University}\\
\email{kleinber@cs.cornell.edu}
}


\maketitle

\begin{abstract}
We study online social networks in which relationships can be
either positive (indicating relations such as friendship) 
or negative (indicating relations such as opposition or antagonism).  
Such a mix of positive 
and negative links arise in a variety of online settings; we study
datasets from Epinions, Slashdot and Wikipedia. We find that
the signs of links in the underlying social networks can be
predicted with high accuracy, using models that
generalize across this diverse range of sites.
These models provide insight into some of the fundamental
principles that drive the formation of signed links in networks,
shedding light on theories of balance and status from
social psychology; they also suggest social computing applications by which 
the attitude of one user toward another can be estimated 
from evidence provided by their relationships with other
members of the surrounding social network.

\end{abstract}

\vspace{1mm} \noindent {\bf Categories and Subject Descriptors:} H.2.8 {\bf
[Database Management]}: Database applications---{\it Data mining}

\vspace{1mm} \noindent {\bf General Terms:} Algorithms; Experimentation.

\vspace{1mm} \noindent {\bf Keywords:} Signed Networks, Structural Balance, Status Theory, Positive Edges,
Negative Edges, Trust, Distrust.

\vspace{-2mm}
\section{Introduction}
\label{sec:intro}
Social interaction on the Web involves both positive and negative relationships
--- people form links to indicate friendship, support, or approval; but they
also link to signify disapproval of others, or to express disagreement or
distrust of the opinions of others.  While the interplay of positive
and negative relations is clearly important in many social network settings, the
vast majority of online social network research has considered only
positive relationships \cite{newman-sirev}.

Recently a number of papers have begun to investigate negative as well as
positive relationships in online contexts. For example, users on Wikipedia can
vote for or against the nomination of others to adminship
\cite{burke-kraut-wikip-promote}; users on Epinions can express trust or
distrust of others \cite{guha-trust,massa05trust}; 
and participants on Slashdot can declare
others to be either ``friends'' or ``foes''
\cite{brzozowski-friend-foe,kunegis-slashdot-zoo,lampe-slashdot}. More
generally, arbitrary hyperlinks on the Web can be used to indicate agreement or
disagreement with the target of the link, though the lack of explicit labeling
in this case makes it more difficult to reliably determine this sentiment
\cite{pang-lee-sentiment-book}.

For a given link in a social network, we will define its {\em sign} to be
positive or negative depending on whether it expresses a positive or negative
attitude from the generator of the link to the recipient.\footnote{We consider
primarily the case of directed links,
though our framework can be applied to undirected links as well.} A fundamental
question is then the following: How does the sign of a given link interact with
the pattern of link signs in its local vicinity, or more broadly throughout the
network?  Moreover, what are the plausible configurations of link signs in real social
networks?  Answers to these questions can help us reason about how negative
relationships are used in online systems, and answers that generalize across 
multiple domains can help to illuminate some of the underlying principles.

Effective answers to such questions can also help inform the design
of social computing applications in which we attempt to infer the
(unobserved) attitude of one user toward another, using the positive and
negative relations that have been observed in the vicinity of this user.
Indeed, a common task in online communities is to suggest new
relationships to a user, by proposing the formation of links to other users
with whom one shares friends, interests, or other properties.
The challenge here is that users may well have pre-existing 
attitudes and opinions --- both positive and negative --- towards others 
with whom they share certain characteristics, and hence before 
arbitrarily making such suggestions to users, it is important
to be able to estimate these attitudes from existing evidence in the network.
For example, if $A$ is known to dislike people that $B$ likes,
this may well provide evidence about $A$'s attitude toward $B$.

\xhdr{Edge Sign Prediction} 
With this in mind, we begin by formulating a concrete underlying task ---
the {\em edge sign prediction problem} ---
for which we can directly evaluate and compare different approaches.
The edge sign prediction problem is defined as follows.
Suppose we are given a
social network with signs on all its edges, but the sign on the edge
from node $u$ to node $v$, denoted $s(u,v)$,
has been ``hidden.'' How reliably can we infer this sign $s(u,v)$ using the
information provided by the rest of the network? 
Note that this problem is both a concrete formulation of
our basic questions about the typical patterns of link signs,
and also a way of approaching our motivating application of inferring
unobserved attitudes among users of social computing sites.
There is an analogy here to 
the {\em link prediction problem} for social networks 
\cite{liben-nowell-link-pred};
in the same way that link prediction is used to
to infer
latent relationships that are present but not recorded by explicit links,
the sign prediction problem can be used to
estimate the sentiment of individuals toward each other, given information
about other sentiments in the network.

In studying the sign prediction problem, we are following an experimental
framework articulated by Guha et al. in their study of trust and distrust on
Epinions \cite{guha-trust}. We extend their approach in a number of directions.
First, where their goal was to evaluate propagation algorithms based on
exponentiating the adjacency matrix, we approach the problem using a
machine-learning framework that enables us to evaluate which of a range of
structural features are most informative for the prediction task. Using this
framework, we also obtain significantly improved performance on the task
itself. Second, we investigate the problem across a range of datasets, and
identify principles that generalize across all of them, suggesting 
certain consistencies in
patterns of positive and negative relationships in online
domains.

Finally, because of the structure of our learned models, we are able to compare
them directly to theories of link signs from social psychology ---
specifically, to theories of {\em balance} and {\em status}. These will be
defined precisely in Section~\ref{sec:signs}, but roughly speaking, balance is
a theory based on the principles that ``the enemy of my friend is my enemy,''
``the friend of my enemy is my enemy,'' and variations on these
\cite{cartwright-harary-balance,heider-balance46}. Status is a theory of signed
link formation based on an implicit ordering of the nodes, in which a positive
$(u,v)$ link indicates that $u$ considers $v$ to have higher status, while a
negative $(u,v)$ link indicates that $u$ considers $v$ to have lower status.
The point is that each of these theories implicitly posits its own model for
sign prediction, which can therefore be compared to our
learned models. The result is both a novel evaluation of these theories on
large-scale online data, and an illumination of our learned models in
terms of where they are consistent or inconsistent with
these theories.

\xhdr{Generalization across Datasets} We study the problem of sign prediction
on three datasets from popular online social media sites; in all cases, we
have network data with explicit link signs. The first is the trust network of
Epinions, in which the sign of the link $(u,v)$ indicates whether $u$ has
expressed trust or distrust of user $v$ (and by extension, the reviews of $v$)
\cite{guha-trust}. The second is the social network of the technology blog
Slashdot, where $u$ can designate $v$ as either a ``friend'' or ``foe'' to
indicate $u$'s approval or disapproval of $v$'s comments
\cite{brzozowski-friend-foe,kunegis-slashdot-zoo,lampe-slashdot}. The third is
the voting network of Wikipedia; here, the sign of the link $(u,v)$ indicates
whether $u$ voted for or against the promotion of $v$ to admin
status \cite{burke-kraut-wikip-promote}.

Despite the fact that link signs have quite different meanings in the
three settings, our main results generalize across all three domains in several
important ways. First, we find that sign prediction performance degrades only
slightly when we train our models on one domain and test them on another. This
indicates that our models are capturing principles that will arguably
generalize to a range of future contexts in which signed links are employed,
rather than picking up on idiosyncrasies of particular individual
domains. Moreover, this generalization holds despite the fact that the quality
of prediction performance is different across the domains: for example,
predicting link signs is more difficult on Wikipedia, yet models trained on
Wikipedia still perform on other domains with very little loss of accuracy
compared to models that were explicitly trained on those domains.

Second, we find that the social-psychological theories of balance and status
agree with the learned models in certain characteristic ways, and disagree in
other characteristic 
ways, as we elaborate in Section~\ref{sec:signs}. These similarities and
contrasts among the models likewise persist at a general level across the
datasets, and thus provide insight into the successes and failures of balance
and status as interpretative frameworks for understanding how link signs are
being used across all these systems.

\xhdr{Additional Tasks} We consider several further issues beyond the problem of
sign prediction. Among these, we ask whether information about negative links
can be helpful in addressing questions that concern purely positive links.
Specifically we consider the {\em link prediction} problem: given a
pair $u$ and $v$, is there a (hidden) positive edge between $u$ and $v$? 
We ask how much performance is improved if the negative edges
in the network are also visible. In other words, how useful is it to know where
a person's enemies are, if we want to predict the presence of additional
friends?  
We find that negative links can be a powerful source of additional information
for a task such as this: 
on all two of the three datasets, we get a boost in improvement
over random choice of up to a factor of 1.5. This type of result helps to argue
that positive and negative links in online systems should be viewed as tightly
related to each other, rather than as distinct non-interacting features of the
system.

We also investigate more ``global'' properties of signed social networks,
motivated by the local theories of balance and status. Specifically, the ``friend of
my enemy'' logic of balance theory suggests that if balance is a key factor in
determining signed
link formation at a global scale, then we should see the network
partition into large opposed factions. The logic of status theory, on the other
hand, suggests that we should see an approximate total ordering of the nodes,
with positive links pointing from left to right and negative links pointing
from right to left. Searching for either of these global patterns involves
developing approximate optimization heuristics for the underlying networks,
since the two patterns correspond roughly to the well-known {\em maximum cut}
and {\em maximum acyclic subgraph} problems. We employ such heuristics, and
find significant evidence for the global total ordering suggested by status
theory, but essentially no evidence for the division into factions suggested by
balance theory. This result provides an intriguing contrast with our basic
results on sign prediction using local features, where strong aspects of both
theories are present; it suggests that the mechanisms by which local organizing
principles scale up to global ones is complex, and an interesting source of
further open questions.

\xhdr{Further Related Work}
Earlier in the introduction, we discussed some of the main lines of research
on which we are building; here, we survey further lines of study
that are also related to our work.

First, our use of trust networks as a source of data connects to 
a large body of work on trust management in several settings, 
including peer-to-peer networks
\cite{kamvar03eigentrust,li04peertrust},
Semantic Web applications \cite{richardson03trust},
and Web spam detection \cite{Gyongyi2004}.
Related to trust management is the development of user
rating mechanisms on sites such as Slashdot
\cite{kunegis-slashdot-zoo,lampe-slashdot} and 
the development of norms to control deviant behavior
\cite{cosley-norms}.
Recent work has also investigated online
communities devoted to discussion of controversial topics,
where one can expect to find strong positive and negative interactions
\cite{brzozowski-friend-foe,wu-huberman-public-opinion}; 
and the analysis of sentiment, subjectivity, and opinion in text
has become an active area in natural language processing
\cite{pang-lee-sentiment-book}.

Our general goal of inferring an individual's attitudes suggests parallels to 
a long line of work on {\em recommendation systems}
\cite{resnick-recommender}, in which the goal is typically
to infer how a user
would evaluate given items based on their evaluation of other items.
There are crucial differences, however, between an analysis in
which a user is evaluating (inert) items, and our case in which a user
is evaluating other {\em people} --- in this latter case, the 
objects being evaluated are themselves capable of 
forming opinions and expressing attitudes, and this provides
additional sources of information based on the full social network
of interactions.

As noted above, there is a long history of work in the social
sciences on balance theory 
\cite{cartwright-harary-balance,heider-balance46},
including more recent work on mathematical models 
that attempt to capture how balance can arise from dynamic
changes to link signs over time
\cite{antal-balance,marvel-balance-landscape}.
In recent work, we analyzed theories of balance and status
in the context of social media sites,
investigating the extent to which each theory helped explain 
the linking behavior of users on these sites \cite{leskovec-submitted}.
Our work there studied how balance and status effects can act
as modifiers on the default behavior of a set of people
measured in aggregate; the problem of making predictions
at the level of individuals was left as an open question.
Here, we take some initial steps toward addressing this question,
combining an analysis of signed networks with machine-learning
techniques so as to formulate individual-level predictions.

\vspace{-2mm}
\section{Dataset Description}
\label{sec:related}
We consider three large online social networks where each link is explicitly
labeled as positive or negative: Epinions, Slashdot and
Wikipedia\footnote{Datasets are available at \url{http://snap.stanford.edu}}.

{{\bf Epinions}} is a product review Web site with a very active user
community. Users are connected into a network of trust and distrust, which is
then combined with review ratings to determine which reviews are most
authoritative. The data spans from the inception of the site in 1999 until
August 12, 2003. The network contains 119,217 nodes and  841,000 edges, of which
85.0\% are positive. 80,668 users received at least one trust or distrust edge,
while there are 49,534 users that created at least one and received at least
one signed edge.

{{\bf Slashdot}} is a technology-related news website. In 2002
Slashdot introduced the {\em Slashdot Zoo}
which allows users to tag each other as
``friends'' or ``foes.''
The semantics of a signed link is similar to Epinions, as a friend
relation means that a user likes another user's comments,
while a foe relationship
means that a user finds another user's comments uninteresting. We crawled
Slashdot in February 2009 to obtain its network of 82,144 users and 549,202
edges of which 77.4\% are positive. 70,284 users received at least one signed
edge, and there are 32,188 users with non-zero in- and out-degree.

{{\bf Wikipedia}} is a collectively authored encyclopedia with an active
user community.  The network we study
corresponds to votes cast by Wikipedia users in
elections for promoting individuals to the role of admin.  A signed link
indicates a positive or negative vote by one user on the promotion of another
($+$ for a supporting vote and $-$ for an opposing vote). Using the latest
complete dump of Wikipedia page edit history (from January 2008) we extracted
all administrator election and vote history data. This gave us 2,794 elections
with 103,747 total votes and 7,118 users participating in the elections (either
casting a vote or being voted on). Out of this total, 1,235 elections resulted
in a successful promotion, while 1,559 elections did not result in the
promotion of the candidate. About half of the votes in the dataset are by the
existing admins, while the other half comes from ordinary Wikipedia users. The
resulting network contains 7,118 nodes and 103,747 edges of which 78.7\% are
positive. There are 2,794 nodes that receive at least one edge
and 1,376 users that both received and created signed edges.

In all networks the background proportion of positive edges is about the same,
with $\approx$80\% of the edges having a positive sign.


\begin{table}[t]
\begin{center}
\small
\begin{tabular}{l||r|r|r}
   & Epinions & Slashdot & Wikipedia \\ \hline \hline
  Nodes & 119,217 & 82,144 & 7,118 \\
  Edges & 841,200 & 549,202 & 103,747 \\
  $+$ edges & 85.0\% & 77.4\% & 78.7\% \\
  $-$ edges & 15.0\% & 22.6\% & 21.2\%
\end{tabular}
\end{center}
\vspace{-3mm} \caption{Dataset statistics.} \label{tab:datasets} \vspace{-3mm}
\end{table}


\vspace{-2mm}
\section{Predicting edge sign}
\label{sec:signs}

We now consider the problem of predicting the sign of individual edges in our
dataset. The set-up for this problem follows the framework of Guha et
al.~\cite{guha-trust}: We are given a full network with all but one of the edge
signs visible, and we are interested in predicting the sign of
this single edge whose sign has been suppressed. This can
be viewed as leave-one-out cross-validation in the present context, where we
learn using the rest of the network and aim to predict the missing sign of a
single edge.

\subsection{A Machine-Learning Formulation}

Given a directed graph $G = (V,E)$ with a sign (positive or negative) on each
edge, we let $s(x,y)$ denote the sign of the edge $(x,y)$ from $x$ to $y$.
That is, $s(x,y)=1$ when the sign of $(x,y)$ is positive, $-1$ when the sign is
negative, and $0$ when there is no directed edge from $x$ to $y$.
Sometimes we will also be interested in the sign of a directed edge
connecting $x$ and $y$, regardless of its direction;
thus, we write $\bar{s}(x,y) = 1$ when there is a positive edge
in one of the two directions $(x,y)$ or $(y,x)$, and either a positive edge
or no edge in the other direction.
We write $\bar{s}(x,y) = -1$ analogously when there is a negative
edge in one of these directions, and either a negative edge or no edge
in the other direction.
We write $\bar{s}(x,y) = 0$ in all other cases (including when
there are edges $(x,y)$ and $(y,x)$ with opposite signs,
though this is in fact rare in our datasets).
For different formulations of our task, we will suppose
that for a particular edge $(u,v)$, the sign $s(u,v)$
or $\bar{s}(u,v)$ is hidden and that we are trying to infer it.

\xhdr{Features} We begin by defining a collection of features for
our initial machine-learning approach to this problem. The features are divided
into two classes. The first class is based on the (signed) degrees of the
nodes, which essentially record the aggregate local relations of a node to the
rest of the world. The second class is based on the principle from social
psychology that we can understand the relationship between individuals $u$ and
$v$ through their joint relationships with third parties $w$: for example,
is there someone who has a positive relationship toward both $u$ and $v$, a
negative relationship toward both $u$ and $v$, or a positive
relationship toward one and a negative relationship toward the other?
Thus, features of this second
class are based on two-step paths involving $u$ and $v$.

We define the first class of features, based on degree, as follows.
As we are interested in predicting the sign of the edge
from $u$ to $v$, we consider outgoing edges from $u$ and incoming edges to $v$.
Specifically we
use $d^+_{in}(v)$ and $d^-_{in}(v)$
to denote the number of incoming positive and negative edges to $v$,
respectively.
Similarly we use $d^+_{out}(u)$ and $d^-_{out}(u)$ to denote
the number of outgoing positive and negative
edges from $u$, respectively.
We use $C(u,v)$ to denote the total number of common neighbors
of $u$ and $v$ in an undirected sense --- that is, the number of nodes $w$ such
that $w$ is linked by an edge in either direction with both $u$ and $v$. We
will also refer to this quantity $C(u,v)$ as the {\em embeddedness} of the edge
$(u,v)$. Our seven {\em degree features} are the five quantities $d^+_{in}(u)$,
$d^-_{in}(v)$,  $d^+_{out}$, $d^-_{out}$, and $C(u,v)$, together with the total
out-degree of $u$ and the total in-degree of $v$, which are $d^+_{out}(u) +
d^-_{out}(u)$ and $d^+_{in}(v) + d^-_{in}(v)$ respectively.

For the second class of feature we consider each {\em triad} involving the edge
$(u,v)$, consisting of a node $w$ such that $w$ has an edge either
to or from $u$ and also an edge either to or from $v$.
There are 16 distinct types of triads involving $(u,v)$: the edge
between $w$ and $u$ can be in either direction and of either sign, and the edge
between $w$ and $v$ can also be in either direction and of either sign; this
leads to $2 \cdot 2 \cdot 2 \cdot 2 = 16$ possibilities. Each of these 16 triad
types may provide different evidence about the sign of the edge from $u$ to
$v$, some favoring a negative sign and some favoring a positive sign. We encode
this information in a 16-dimensional vector specifying the number of triads of
each type that $(u,v)$ is involved in.

\begin{figure}[!t]
  \begin{center}
    \includegraphics[width=0.4\textwidth]{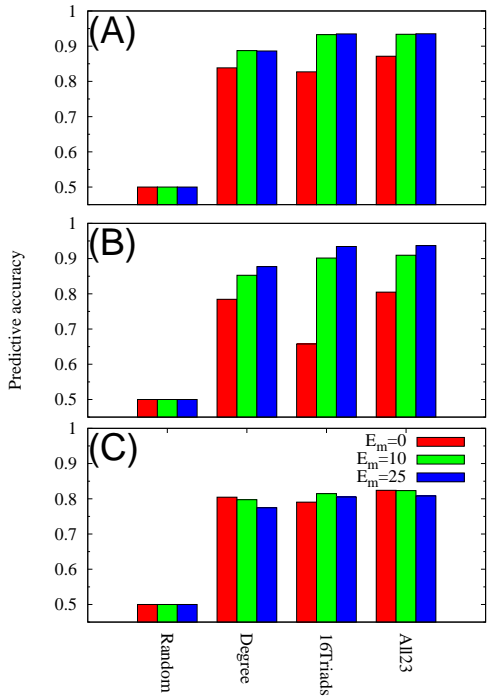}
    \vspace{-9mm}
    \caption{Accuracy of predicting a sign of edge $(u,v)$ given signs of all
    other edges in the network. (a) Epinions, (b) Slashdot, (c) Wikipedia.}
    \vspace{-9mm}
    \label{fig:edgePred1}
  \end{center}
\end{figure}

\xhdr{Learning Methodology and Results} We use a logistic regression classifier
to combine the evidence from these individual features into an edge sign
prediction. Logistic regression learns a model of the form
$$
P(+|x) = \frac{1}{1+e^{- (b_0 + \sum_i^n b_i x_i})}
$$
where $x$ is a vector of features ($x_1, \dots, x_n$) and $b_0, \dots, b_n$ are
the coefficients we estimate based on the training data.

The edges signs in the networks that we study are overwhelmingly positive. Thus
we consider and evaluate two different approaches.
First, we use the full dataset where about 80\% of the
edges are positive. Second, we follow the methodology of Guha et al.
\cite{guha-trust} and create a {\em balanced dataset}
with equal numbers of positive and
negative edges, so that random guessing yields a 50\% correct prediction rate.
For every negative edge $(u,v)$ we sample a random positive edge, which ensures
that the number of positive and negative edges in the data we consider for
training and prediction is balanced. Moreover, we also consider two different
evaluation measures: the classification accuracy and the area under the ROC
curve (AUC). For ease of exposition we focus on classification accuracy on a
balanced dataset. As we discuss later our results are robust to whether we
use the full or balanced dataset and whether we evaluate using AUC or
accuracy.

We describe each edge $(u,v)$ in this set using the two classes of features
described above. We consider all 23 features together, and we also evaluate
performance using features of each class separately --- that is, representing
each edge as a 7-dimensional vector of degree features and as a 16-dimensional
vector of triad features. We also consider performance across different types
of edges. In particular, since the triad features are relevant only when $u$
and $v$ have neighbors in common, it is natural to expect that they will be
most effective for edges of greater embeddedness. We therefore consider the
performance restricted to subsets of edges of different levels of minimum
embeddedness.

The classification accuracy is shown in Figure~\ref{fig:edgePred1}, where
results are described for all three datasets, for the two classes of features
separately and together, and for different levels of minimum embeddedness
(denoted by $E_m$). Several observations stand out. First, prediction based on
the learned models significantly outperform the results reported in Guha et
al.~\cite{guha-trust} for the Epinions dataset. The lowest error rate achieved
in their paper is 14.7\% whereas we obtain error rates of 11.45\% for Degree,
6.64\% for 16Triads and 6.58\% for All23.

These results are particularly interesting because our features are based only
on local properties in the one-step neighborhood of the edge $(u,v)$ whose sign
is being inferred, in contrast with the propagation model of Guha et al.  This
suggests that edge signs can be meaningfully understood in terms of such local
properties, rather than requiring a notion of propagation from farther-off
parts of the network.

Second, consistent with intuition, the triad features perform less well than
the degree features for edges of low embeddedness, but become more effective as
the embeddedness increases and a greater amount of triadic information becomes
available.

Finally, it is also noteworthy that the accuracy on the Wikipedia network is
significantly lower than on the other two networks, even for edges
with large embeddedness. This discrepancy between Wikipedia and the other
datasets is interesting because the positive and negative links on Wikipedia
correspond to evaluations that are more publicly visible, more consequential,
and more information-based than for the other two datasets, since they result
from public votes on promotion of individuals to adminship, where the
candidates being voted on have accumulated a long history of activity on
Wikipedia. One could conjecture that these aspects of the evaluations in the
Wikipedia dataset make it correspondingly more difficult (though still
surprisingly feasible) to predict their outcomes from simple structural
measures.

In all experiments we report the average accuracy and estimated logistic
regression coefficients over 10-fold cross validation. If not stated otherwise,
we limit our analyses to edges with minimum embeddedness 25. We note that our
results are robust with respect to training dataset and evaluation metric.
Generally, when using the full dataset rather than the balanced one,
random guessing improves accuracy from 50\%
to approximately 80\%. With the full dataset
the accuracy of our logistic regression method correspondingly jumps
to the 90-95\% range and maintains roughly a 15\% absolute improvement over
random guessing.
When evaluating using AUC rather than accuracy
the overall pattern of performance does not change.
The various forms of logistic regression have AUC of approximately
90\% on the balanced dataset and 95\% on the full dataset.

\subsection{Connections to Theories of Balance and Status}

Our goal is to use the machine learning framework not just to predict the edge
signs themselves, but also for deriving insights into the usage of these
systems based on the observed patterns of positive and negative
edges.

Specifically, logistic regression provides a coefficient associated with each
feature, which suggests how the feature is being used by the model to provide
weight for or against a positive edge sign. This provides a natural and
appealing connection to classical theories from social psychology, which also
offer proposals for how subsets of these features offer evidence for the sign
of the edge $(u,v)$.

We focus here on the second class of features, based on triad types, which are
motivated by social-psychological theories about local patterns of
relationships. Specifically,
we say that a {\em theory of triad types} is a function
$$f : \{{\rm types} ~ \tau\} \rightarrow \{+1,-1,0\},$$
which specifies for each triad type $\tau$ whether it constitutes evidence for
a positive $(u,v)$ edge ($f(\tau) = +1$), evidence for a negative $(u,v)$ edge
($f(\tau) = -1$), or whether it offers no evidence ($f(\tau) = 0$).

Our logistic regression model provides a learned theory of triad types for each
dataset, in which $f(\tau)$ is equal to the sign of the coefficient associated
with the feature $\tau$. But several principles from social psychology also
provide theories of triad types, developed from plausible assumptions about
human behavior rather than through a data-driven approach. In other words, the
learned model and the qualitative models from the literature are expressed in
the same language --- as mappings from triad types to positive or negative
evidence --- and we can thus ask questions about how the theories align with
each other. Through this line of investigation we can gain insight
into two issues: first,
we can evaluate the existing theories on our on-line datasets; and second, we
can use these existing theories as an interpretive framework for reasoning
about the structure of our learned model.

\xhdr{Balance and Status} We begin by summarizing the two main
social-psychological theories of triad types that we compare to.

The more well-studied of the two
is {\em structural balance theory}, based on
the common principles that
``the friend of my friend is my friend,''
``the enemy of my friend is my enemy,''
``the friend of my enemy is my enemy,''
and (perhaps less convincingly)
``the enemy of my enemy is my friend.''
Concretely, this means that if $w$ forms a triad with the edge
$(u,v)$, then structural balance theory posits that $(u,v)$
should have the sign that causes the triangle on $u, v, w$ to
have an odd number of positive signs, regardless of edge
direction --- just as each of the
principles above has an odd number of occurrences of the word ``friend.''
In other words, $f_{balance}(\tau) = \bar{s}(u,w) \bar{s}(v,w)$, where
we recall that the value of $\bar{s}$ corresponds to the sign
regardless of the direction of the edge.

An alternate theory, which is implicit in the work of
Guha et al. \cite{guha-trust} and developed further
in our recent research \cite{leskovec-submitted}, is a theory of {\em status}.
In this theory, a positive edge $(x,y)$ means that $x$ regards $y$ as having
higher status than herself, while a negative edge $(x,y)$ means that $x$
regards $y$ as having lower status than herself. Assuming that all participants
in the system agree on this status ordering, status theory predicts that when
the direction of an edge is flipped, its sign should flip as well.

So to determine $f_{status}(\tau)$, we first flip the directions of the edges
between $u$ and $w$ and between $v$ and $w$, if necessary, so that they point
from $u$ to $w$ and from $w$ to $v$; we flip the signs accordingly as we do
this. We then define $f_{status}(\tau)$ to be the sign of $s(u,w) + s(w,v)$.
This means that status theory makes no prediction when the two signs cancel
out, but otherwise, it predicts a positive or negative sign based on the
imputed status relationship between $u$ and $v$.

Notice that balance and status agree on some types of triads --- for example,
when $u$ points positively to $w$ and $w$ points positively to $v$, then $v$ is
both the friend of $u$'s friend, and also someone of higher status than $u$,
and
thus both theories predict a positive sign for $(u,v)$. But balance and status
can also disagree --- for example, when $v$ points positively to $w$ and $w$
points positively to $u$, then balance concludes that $v$ is the friend of
$u$'s friend and thus $(u,v)$ is positive, but status posits that $v$ has lower
status than $u$ and thus $(u,v)$ is negative.

\begin{table*}[!t]
  \small
  \begin{center}
  \begin{tabular}{l||r|r||r|r|r||r|r|r||r|r|r}
  Feature & Bal & Stat & \textbf{Epinions} & Bal & Stat & \textbf{Slashdot} & Bal & Stat & \textbf{Wikipedia} & Bal & Stat \\ \hline \hline
  const &   &   & -0.1656 &  &  & 0.018 &  &  & -0.215 &  & \\
  FFpp & 1 & 1 & 0.4869 &   &   & 0.8504 &   &   & 0.2849 &   & \\
  FFpm & -1 & 0 & -0.5166 &   &   & -0.9008 &   &   & -0.4337 &   & \\
  FFmp & -1 & 0  & -0.4476 &   &   & -1.0513 &   &   & -0.3092 &   & \\
  FFmm & 1 & -1 & -0.7331 &$\times$&   & -0.5874 &$\times$&   & -0.768 &$\times$& \\
  FBpp & 1 &  0 & 0.3416 &   &   & 0.4385 &   &   & 0.0544 &   & \\
  FBpm & -1 & 1 & -0.0147 &   &$\times$& -0.1439 &   &$\times$& -0.0131 &   &$\times$\\
  FBmp & -1 & -1 & -0.8598 &   &   & -1.1887 &   &   & -0.1986 &   & \\
  FBmm & 1 & 0  & 0.0436 &   &   & -0.0719 &$\times$&   & -0.0325 &$\times$& \\
  BFpp & 1 &  0 & 0.0814 &   &   & 0.3593 &   &   & 0.116 &   & \\
  BFpm & -1 & -1 & -1.3097 &   &   & -1.0838 &   &   & -0.3527 &   & \\
  BFmp & -1 & 1 & -0.1228 &   &$\times$& -0.248 &   &$\times$& 0.0527 &$\times$& \\
  BFmm & 1 &  0 & 0.0788 &   &   & -0.024 &$\times$&   & -0.0968 &$\times$& \\
  BBpp & 1 & -1 & -0.0855 &$\times$&   & -0.0873 &$\times$&   & -0.0065 &$\times$& \\
  BBpm & -1 & 0  & -0.0536 &   &   & -0.2736 &   &   & -0.0168 &   & \\
  BBmp & -1 & 0 & -0.0382 &   &   & -0.2788 &   &   & 0.0507 &$\times$& \\
  BBmm & 1 & 1 & -0.0242 &$\times$&$\times$& 0.2275 &   &   & -0.1616 &$\times$&$\times$\\ \hline
  \multicolumn{3}{l||}{Total errors}  &  & 3 & 3 &  & 4 & 2 &  & 7 & 2
  \end{tabular}
  \end{center}
    \vspace{-6mm}
  \caption{Logistic regression coefficients compared to status and structural
  balance theory. $\times$ means there is discrepancy in predictions between
  the Balance (Status) theory and what is learned from the logistic regression
  model. Each line represents directions
  and signs of the edges on a path $(A,B,C)$ where ``BFpm" stands for
  \textbf{B}ackward \textbf{F}orward \textbf{p}lus \textbf{m}inus and denotes a path
  $A \leftarrow_{+} B \rightarrow{-} C$.
  }
  \label{tab:pred}
    \vspace{-2mm}
\end{table*}

\xhdr{Comparison of Balance and Status with the Learned Model} In
Table~\ref{tab:pred}, we show the signs of the three theories discussed
above --- balance, status, and the learned model --- on the three datasets. For
denoting the 16 triad types, in the table and elsewhere, we use a shorthand in
which we record the four binary choices that comprise each type. Thus a type
$\tau$ will be represented by a string of the form $[F|B][F|B][p|m][p|m]$ to
indicate the direction of the edges along the two-step path $u$-$w$-$v$
(Forward or Backward on each step), and the signs of these two edges (plus or
minus). For example, $FBmp$ is the triad type in which $u$ points negatively to
$w$, and $v$ points positively to $w$ (since the first step in the $u$-$w$-$v$
path is forward and minus, while the second is backward and plus).

At a general level, the results show that both social-psychological theories
agree fairly well with the learned models --- with agreement on more than half
the triad types where they make predictions, and generally on
three-quarters or more of the triad types.
Looking at the absolute values of the coefficients, we note that
certain features stand out in importance across all three datasets ---
specifically, there are coefficients of large magnitude for
all the FF features, as well as the FBmp and BFpm features.
In contrast, the BB features have coefficients of much smaller magnitude.
We also see that balance theory is
in notably better alignment with the learned model for Epinions and Slashdot
than it is for Wikipedia. As discussed above, Wikipedia differs considerably
from the other two datasets in that it is a publicly visible voting forum.
Given these results it is interesting to conjecture that in such a setting
status may play a stronger role.

It is also interesting to consider the cases in which there are relatively
stable disagreements among the models across the three datasets.  In particular,
we see that balance theory consistently disagrees with the learned model (and
with status theory) when it predicts that a negative $(u,w)$ and negative
$(w,v)$ edge should suggest a positive $(u,v)$ edge.  This is precisely the
kind of case that seems somewhat suspect intuitively, namely ``the enemy of my
enemy is my friend''. Balance theory also consistently disagrees with both the
learned model and status theory when it predicts that a positive $(v,w)$ and
positive $(w,u)$ edge should result in a positive $(u,v)$ edge.  Here the
direction of the two-step path is from $v$ to $u$ rather than $u$ to $v$, and
one can conjecture that this opposite direction path has a lower predictive
power for the $(u,v)$ sign.   Indeed these two cases suggest modifications of
the models, as we now discuss.

\begin{table}[t]
  \small
  \begin{center}
  \begin{tabular}{l||r|r|r|r}
  Feature & Balance theory & Epinions & Slashdot & Wikipedia \\ \hline \hline
  const & 0 & 0.4321 & 1.4973 & 0.0395 \\
  pp & 1 & 0.0470 & 0.0395 & 0.0553 \\
  pm & -1 & -0.1154 & -0.2464 & -0.1632 \\
  mp & -1 & -0.2125 & -0.3476 & -0.1432 \\
  mm & 1 & -0.0149 & -0.0262 & -0.0465
  \end{tabular}
  \end{center}
  \vspace{-3mm}
  \caption{Regression coefficients based on Balance attributes
  and learned logistic regression.}
  \vspace{-3mm}
  \label{tab:balancePred}
\end{table}

\begin{table}[t]
  \small
  \begin{center}
  \begin{tabular}{l||r|r|r|r}
  Feature & Status theory & Epinions & Slashdot & Wikipedia \\ \hline \hline
  const & 0 & -0.6873 & -1.3915 & -0.3039 \\
  $u<w<v$ & 1 & 0.1165 & 0.0463 & 0.0258 \\
  $u>w>v$ & -1 & -0.1002 & -0.114 & -0.1941 \\
  $u<w>v$ & 0 & 0.0572 & 0.1558 & 0.0300 \\
  $u>w<v$ & 0 & -0.0064 & 0.0382 & 0.0543
  \end{tabular}
  \end{center}
  \vspace{-3mm}
  \caption{The coefficients based on Status Theory and
 learned logistic regression.}
 \vspace{-5mm}
  \label{tab:statusPred}
\end{table}

\begin{table}[t]
  \small
  \begin{center}
  \begin{tabular}{l||r|r|r}
    Feature & Epinions & Slashdot & Wikipedia \\ \hline \hline
    const & -2.2152 & -2.8475 & -1.4496 \\
    FF & 0.2023 & 0.2092 & 0.0773 \\
    FB & 0.1286 & 0.1698 & 0.0286 \\
    BF & 0.0077 & 0.0842 & 0.0544 \\
    BB & -0.0692 & -0.0293 & -0.0259 \\
  \end{tabular}
  \end{center}
  \vspace{-4mm}
  \caption{Learned logistic regression coefficients for the model
  based on the counts of directed positive paths.}
  \label{tab:plusPathsPred}
  \vspace{-5mm}
\end{table}

\xhdr{Comparison of Balance and Status with Reduced Models} To fully understand
the relationship of the learned model to the theories of balance and status, it
helps to look at ``reductions'' of the model that capture just the features
essential to these two theories.

Let's begin by considering balance theory. Balance theory has generally been
applied as a theory of undirected graphs, although its extension to directed
graphs by ignoring edge direction (as we use it here) is standard as well
\cite{wasserman-faust}. With this in mind, let's consider the learning problem
using a feature set in which we treat all edges as undirected. In this setting,
there are only four different triad types involving a node $w$ and the edge
$(u,v)$, depending on whether the undirected edge $\{u,w\}$ is positive or
negative, and whether the undirected edge $\{w,v\}$ is positive or negative
(since we can no longer observe the directions of these edges). Thus, we create
a 4-dimensional feature vector for the edge $(u,v)$, by simply counting how
many undirected triads of each type it is involved in. We then apply logistic
regression to this 4-dimensional problem.

The results are depicted in Table~\ref{tab:balancePred}. We see that for all
triad types other than the ``enemy of my enemy'' type ($mm$), and all three
datasets, the learned coefficient is the same as the prediction of balance
theory. The disagreement for the $mm$ type is a further indication of the
difficulty of the ``enemy of my enemy'' aspect of balance in these domains, and
in fact is consistent with an alternative formulation of balance theory due to
Davis in the 1960s \cite{davis-balance}, which agrees with standard balance
theory on the first three triad types and makes no prediction on $mm$.
We will refer to Davis's variant on balance theory as {\em weak balance}.

We can do a similar reduction for status theory. We begin by preprocessing the
graph to flip the direction and sign of each negative edge, thereby creating a
positive edge with the same interpretation under status theory. The resulting
graph has only positive edges, and hence there are only four triad types ---
based on whether the $(u,w)$ edge is forward or backward, and whether the
$(w,v)$ edge is forward or backward. We create a 4-dimensional feature vector
for the edge $(u,v)$ by counting the frequencies of these four triad types, and
apply logistic regression.

The learned coefficients for this problem are shown in
Table~\ref{tab:statusPred}. Here we see that on both triad types for which
status theory makes a prediction, and across all three datasets, the sign of
the learned coefficient is the same as the sign of the status prediction.

What emerges from the analysis of these reduced-form models is that each of
balance theory and status theory are essentially accurate at the granularity in
which they are most naturally formulated --- in the case of balance, on
undirected graphs; and in the case of status, once edge signs and directions
have been canonicalized. This makes it clear that our results for the more
detailed 16-type model go beyond the scale of resolution at which either
balance or status can provide accurate predictions, and illuminate some more
subtle effects that govern social interactions.

\begin{figure}[!t]
  \begin{center}
    \vspace{-1mm}
    \includegraphics[width=0.4\textwidth]{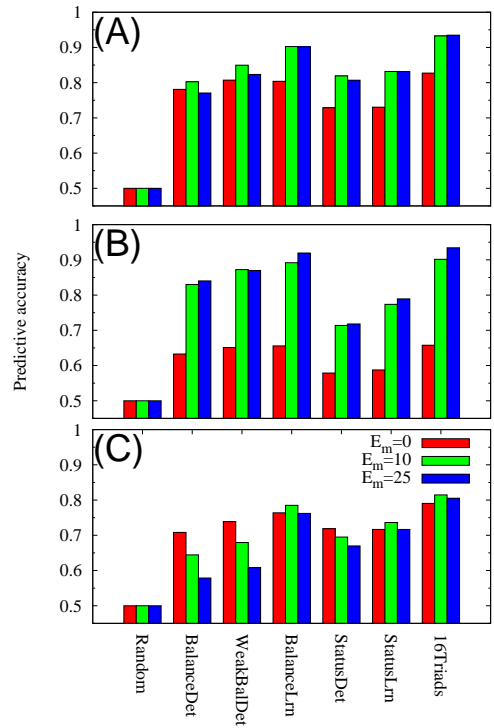}
    \vspace{-4mm}
    \caption{Accuracy of predicting a sign of edge $(u,v)$ given signs of all
    other edges in the network. (a) Epinions, (b) Slashdot, (c) Wikipedia.}
    \vspace{-8mm}
    \label{fig:edgePred2}
  \end{center}
\end{figure}

To further elaborate on this point, we can evaluate the prediction
accuracy of each of these reduced models in comparison to
the full model on all 16 triad types.
The results are shown in Figure~\ref{fig:edgePred2}.
Here, we consider the two kinds of reduced 4-dimensional feature sets,
and evaluate performance using both the coefficients learned
via logistic regression
(denoted {\em BalanceLrn} and {\em StatusLrn} in the figure),
as well as the lower performance using coefficients from $\{-1, 0, +1\}$
provided by balance, Davis's notion of weak balance, and status
(denoted {\em BalanceDet}, {\em WeakBalDet},
and {\em StatusDet} in the figure).

\xhdr{All-positive subgraphs}
There is a final reduced model that also provides insight
into the role of balance theory particularly for these datasets.
Suppose that we preprocess the graph by simply deleting all negative
edges, so that we are left with the subgraph consisting
of only positive edges.
Again, there are now four possible triad types, and we show the
learned coefficients for logistic regression on this 4-dimensional problem
in Table~\ref{tab:plusPathsPred}.
Balance would predict that all coefficients should be positive, since
all relations indicate friendship under the interpretation of balance theory.
This agrees with Table~\ref{tab:plusPathsPred} except for the last
row, where the coefficient of the learned model is negative across
all three datasets.
This corresponds to the triad type in which $v$ links positively to $w$,
and $w$ links positively to $u$.
The data indicates that this in fact serves as evidence of a
negative $(u,v)$ link in all datasets, and status theory provides one
simple hypothesis for why: if $v$ regards $w$ as having higher status,
and $w$ in turn regards $u$ as having higher status, then arguably
$u$ will view $v$ as having lower status.

\begin{figure*}[t]
  \begin{center}
  \begin{tabular}{ccc}
    \includegraphics[width=0.32\textwidth]{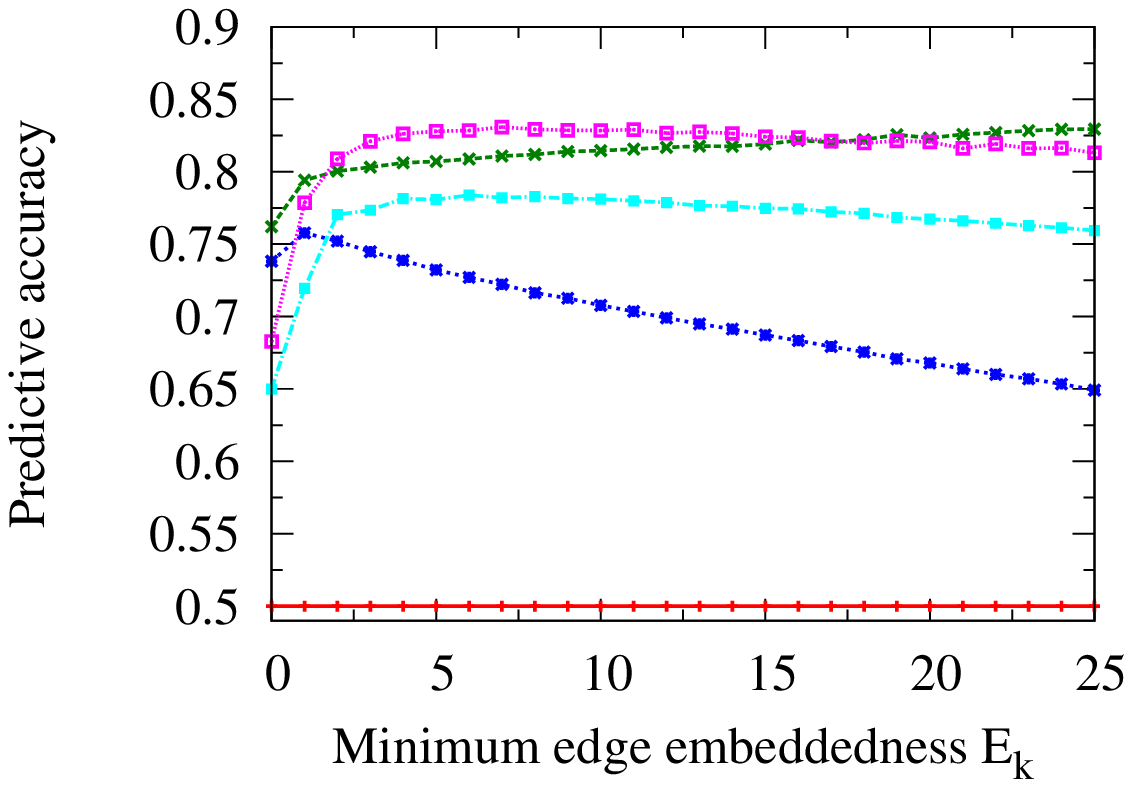} &
    \includegraphics[width=0.32\textwidth]{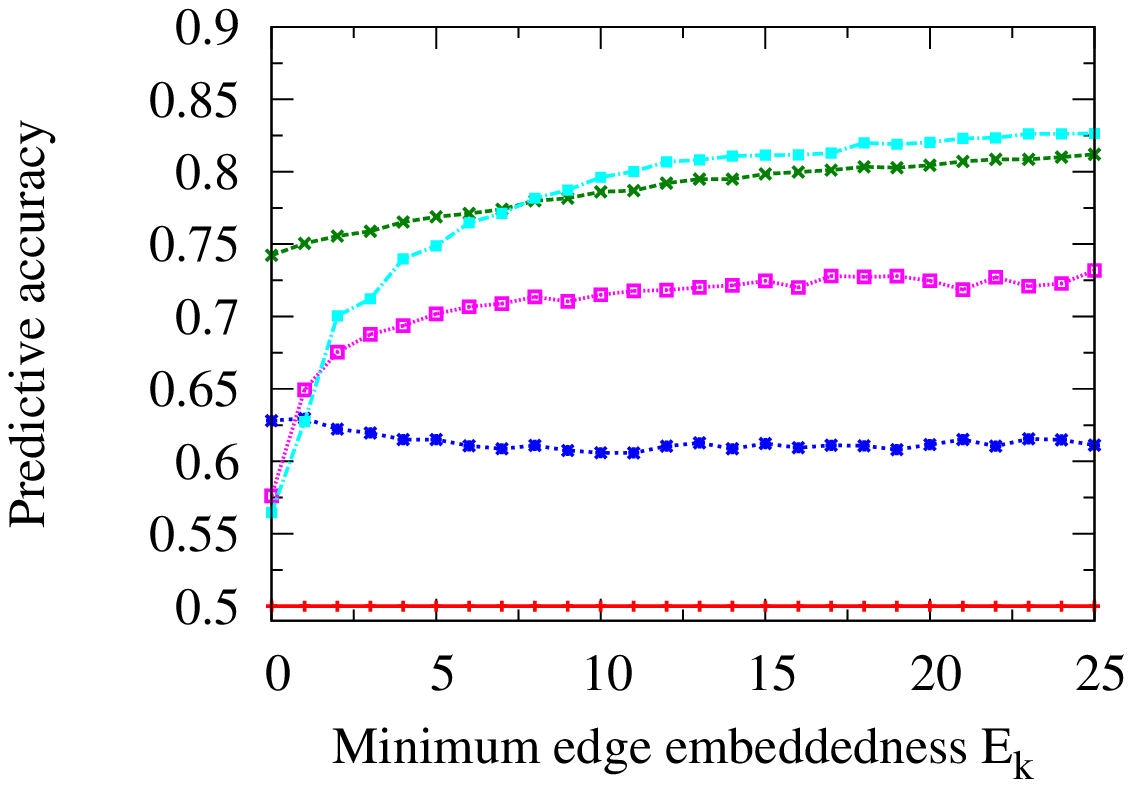} &
    \includegraphics[width=0.32\textwidth]{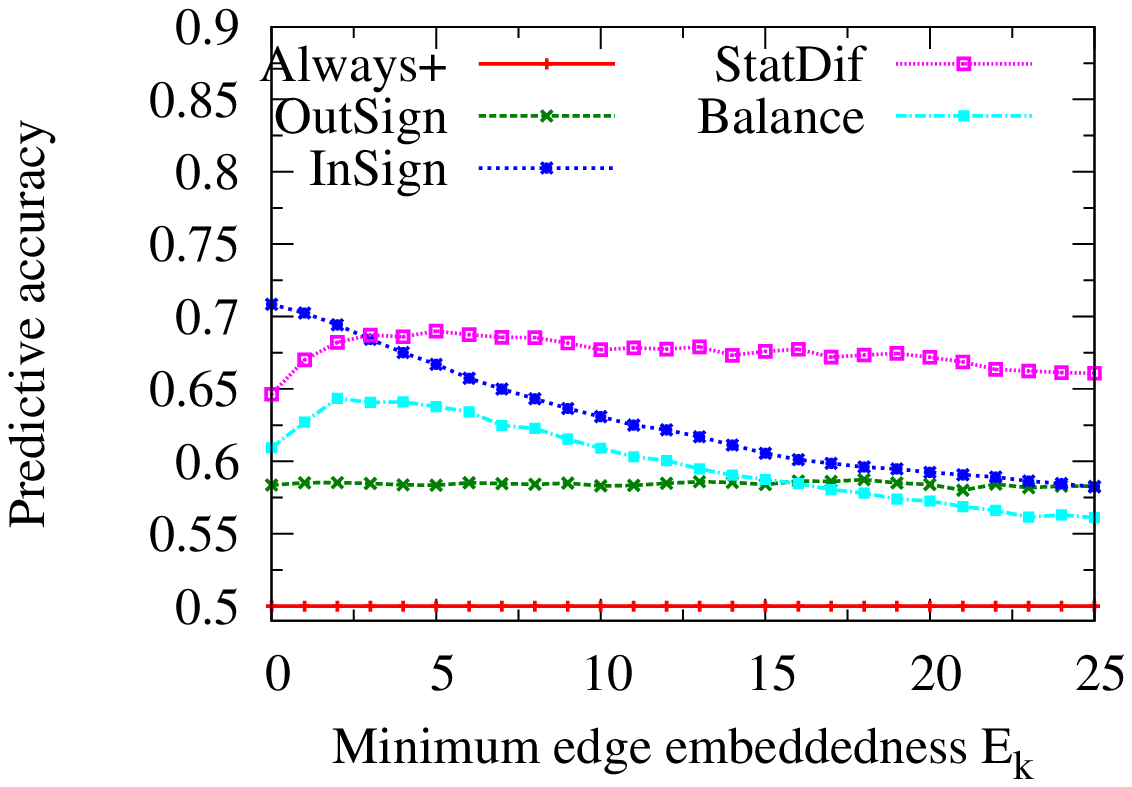} \\
    (a) Epinions & (b) Slashdot & (c) Wikipedia
  \end{tabular}
  \vspace{-3mm}
  \caption{Accuracy for simple models as a function of
  minimum edge embeddedness. Refer to main text for model descriptions.}
  \label{fig:edgePred}
    \vspace{-6mm}
  \end{center}
\end{figure*}

\subsection{Generalization across datasets}

We now turn to the question of how well the learned predictors
generalize across the three datasets, in order to investigate the
extent to which the learned models are based on specific properties of
each dataset versus more general properties of social network data.
That is, in our investigation thus far the learned models have been
able to take advantage of properties of each specific dataset, whereas
the base social science models are generic across datasets.
Thus it could be the case that the models obtained using logistic regression
perform well only on the individual datasets on which they were
trained.  On the other hand, if the learned models are able to
generalize across datasets it suggests that
there are underlying general principles that guide the creation of signed
edges in social network graphs and which the regression models
are able to capture.

\begin{table}[t]
  \small
  \begin{center}
  \begin{tabular}{l||r|r|r}
    {\bf All23} & Epinions & Slashdot & Wikipedia  \\ \hline \hline
    Epinions & 0.9342 & 0.9289 & 0.7722 \\
    Slashdot & 0.9249 & 0.9351 & 0.7717 \\
    Wikipedia & 0.9272 & 0.9260 & 0.8021
  \end{tabular}
  \vspace{-3mm}
  \caption{Predictive accuracy when training on the ``row'' dataset and
  evaluating the prediction on the ``column'' dataset.}
  \vspace{-6mm}
  \label{fig:generaliz}
  \end{center}
\end{table}

To evaluate the generalization accuracy of the models we perform the
following experiment.  For each pair of datasets we train the
logistic regression model on the first dataset and evaluate it on the second
dataset.
Table~\ref{fig:generaliz} shows the results of these 9 experiments using the
All23 model.  The diagonal entries of the table show the results already
presented above (i.e. leave-one-out cross-validation for a single dataset),
whereas the off-diagonal entries show the generalization across datasets.  We
see the same overall pattern as before, with prediction accuracy being
considerably lower for Wikipedia than for the other two dataset. We also see
that the off-diagonal entries are nearly as high as the diagonals, showing that
there is very good generalization and thus there indeed are general
cross-dataset properties captured by the model. In both the first and second
columns (testing on the Epinions and Slashdot datasets) there is remarkably
little decrease in performance regardless of which of the three datasets is
used for training.  Note that in particular, even training on the Wikipedia
dataset yields good prediction performance on the other two datasets -- thus
while the Wikipedia dataset is difficult in terms of prediction accuracy it
seems to provide the same underlying structural information as the other
datasets in that it allows training of a more general model.

\begin{table}[t]
  \small
  \begin{center}
  \begin{tabular}{l||r|r|r}
    {\bf BalanceLrn} & Epinions & Slashdot & Wikipedia \\ \hline \hline
    Epinions & 0.9027 & 0.9166 & 0.7319 \\
    Slashdot & 0.9020 & 0.9203 & 0.7439 \\
    Wikipedia & 0.8985 & 0.9145 & 0.7558 \\ \hline
    BalanceDet  & 0.7709 & 0.8414 & 0.5778  \\
    WeakBalance & 0.8233 & 0.8698 & 0.6081  \\
    \multicolumn{4}{c}{}\\
    {\bf StatusLrn} & Epinions & Slashdot & Wikipedia \\ \hline \hline
    Epinions & 0.8313 & 0.7514 & 0.6410 \\
    Slashdot & 0.7682 & 0.7847 & 0.6094  \\
    Wikipedia & 0.7592 & 0.6598 & 0.7163  \\ \hline
    StatusDet & 0.8083 & 0.7173 & 0.6679
  \end{tabular}
  \vspace{-2mm}
  \caption{Predictive accuracy when training on the ``row'' dataset and
  evaluating the prediction on the ``column'' dataset.}
  \label{fig:generaliz2}
  \vspace{-6mm}
  \end{center}
\end{table}

Table~\ref{fig:generaliz2} shows the results of these experiments
using the learned Balance and Status models considered above.  For
comparison we also show the performance of the basic BalanceDet, WeakBalance
and StatusDet models (which are not learned from the data).  For balance we
see that the generalization performance is again very good, and also
that the prediction accuracy in all cases is higher than for the
nonlearned balance models.  However for status, we see a considerably different
picture. The model does not generalize as well, and in fact often
performs worse than the baseline status model when trained on one
dataset and tested on another.  This suggests that the learned balance
properties are relatively generic across different online settings,
whereas the status properties are more particular to a given dataset.

\subsection{Heuristic Predictors}

Having now explored the relationship of balance and status to the output of a
machine learning model, it is natural to also investigate simple
``hand-written'' heuristic predictors to determine baseline performance levels
for the sign prediction task.

We construct a few such predictors using ideas drawn from status and balance
theory, as well as considerations of node degree. We again use the same
evaluation framework, in which for every negative edge $(u,v)$ we sample a
random positive edge, to ensure that random guessing gets 50\% of the
predictions correct.

We consider the following heuristic predictors:
\begin{itemize}
\denselist
\item {\em A balance heuristic.}  For each choice of the sign of $(u,v)$,
    some of the triads it participates in will be consistent with balance
    theory, and the rest of the triads will not. We choose the sign for
    $(u,v)$ that causes it to participate in a greater number of triads
    that are consistent with balance.
\item {\em A status heuristic.} We define an estimate of a node $x$'s
    status to be $\sigma(x) = d^+_{in}(x) + d^-_{out}(x) - d^+_{out}(x) -
    d^-_{in}(x).$ This gives $x$ status benefits for each positive link it
    receives and each negative link it generates, and status detriments for
    each negative link it receives and each positive link it generates. We
    then predict a positive sign for $(u,v)$ if $\sigma(u) \leq \sigma(v)$,
    and a negative sign otherwise.
\item {\em An out-degree heuristic.} We predict the majority sign based on
    the signs given by the edge initiator $u$. That is, we predict $+$ if
    $d^+_{out}(u)\ge d^-_{out}(u)$. \item {\em An in-degree heuristic.} We
    predict the majority sign based on the signs received by the edge
    target $v$. That is, we predict $+$ if $d^+_{in}(v)\ge d^-_{in}(v)$.
\end{itemize}

We show the results of these simple predictors in Figure~\ref{fig:edgePred},
plotted as a function of embeddedness. First, while these predictors perform
considerably better than guessing -- and quite well in some cases -- they
generally perform worse than the models discussed above.  Second, in
the case of the Epinions data, some of these simple predictors perform
comparably well to the results reported in \cite{guha-trust} (though not as
well as the very best results in that paper).  This underscores the value of
having simple baseline measures.

There are several additional points worth noting.  First, the
in-degree heuristic (InSign) performs relatively poorly across all
datasets (although it beats the other predictors for low embeddeness
in the Wikipedia dataset), while out-degree (OutSign) performs well on
two of the three; this corresponds to the intuitively natural
conclusion that the generator of the edge plays a larger role in sign
determination than the recipient.  Second, the triadic features ---
status (StatDif) and balance (Balance) --- do better with increasing
embeddedness, but in most cases performance of these features starts
to drop again as embeddedness gets too high. One can understand this
in terms of a tradeoff between two types of information sparsity.  On
the one hand, when an edge participates in relatively few triads (low embedding)
then the triadic features provide relatively little information, but
on the other hand
relatively few edges in the graph have a high
degree of embedding (e.g., significantly above 25).


\vspace{-2mm}
\section{Global structure of signed networks}
\label{sec:global}


When we perform sign prediction for an edge $(u,v)$ using information about the
two-step paths from $u$ to $v$, and when we relate our learned models to the
predictions of balance and status, we are using {\em local} information about
the neighborhoods of the nodes. However, the theories of balance and status
also each make {\em global} predictions about the pattern of the signs in the
network as a whole, and it is interesting to investigate the extent to which
these global predictions are borne out by the network structure of our
datasets.

\xhdr{Balance and Status: From Local to Global}
The global predictions of balance and status are best explained
in their simplest settings, which are for networks in which
each pair of nodes is connected by an edge.
This complete connectivity clearly does not hold for our datasets,
but we will explain how to adapt the predictions of the models
to our setting nonetheless, and then search for evidence of
these predictions.

The global prediction of balance theory is contained in
a theorem of Cartwright and Harary from the 1950s.
It asserts that if all triangles in a completely connected
undirected graph obey structural balance, then globally the network can
be divided into two cliques of mutual friends, such that all
edges between the two cliques are negative
\cite{cartwright-harary-balance}.
\begin{theorem}[Cartwright-Harary]
Let $G$ be a signed, undirected complete graph in which each
triangle has an odd number of positive edges.
Then the nodes of $G$ can be partitioned into two sets $A$ and $B$
(where one of $A$ or $B$ may be empty),
such that all edges within $A$ and $B$ are positive, and
all edges with one end in $A$ and the other in $B$ are negative.
\label{thm:balance}
\end{theorem}

We can analogously formulate a local-to-global connection for status theory,
but it leads to a quite different structural prediction.
Rather than undirected complete graphs, the basic form of this
theorem will use completely connected directed graphs, which are
referred to as {\em tournaments}: directed graphs in which
each pair of nodes is connected by a directed edge
in one direction or the other.

First, if we consider the local condition that motivated status
theory in the previous section, it required that for any edge
$(u,v)$, and any third node $w$, it should be possible to
assign distinct numerical
``status values'' to $u$, $v$, and $w$ in such a way that
the positive edges among them (if any) go from
nodes of lower status to nodes of higher status, and the
negative edges among them (if any) go
from nodes of higher status to nodes of lower status.
Let us say that the three nodes $u$, $v$, and $w$ are
{\em status-consistent} if this condition holds.
We can now have the following result, which says that
if all three-node sets are status-consistent, then the whole
graph obeys a form of status-consistency: there is
a total ordering of the nodes in which positive edges all go left-to-right,
and negative edges all go right-to-left.
\begin{theorem}
Let $G$ be a signed, directed tournament, and
suppose that all sets of three nodes in $G$ are status-consistent.
Then it possible to order the nodes of $G$ as $v_1, v_2, \ldots, v_n$
in such a way that each positive edge $(v_i,v_j)$ satisfies $i < j$,
and each negative edge $(v_i,v_j)$ satisfies $i > j$.
\label{thm:status}
\end{theorem}
\xproof{
Following an idea from the previous section, we first reverse the
direction of each negative edge in $G$ and give it a positive sign.
Notice that if all sets of three nodes were status-consistent before this
conversion, they remain status-consistent after this conversion.

Let $G'$ be the resulting graph; note that $G'$ has only positive edges.
If any three-node subgraph in $G'$ were to form a directed cycle,
then the three nodes in this cycle would violate status-consistency.
Thus, all three-node subgraphs of $G'$ are acyclic.
Applying a well-known theorem on tournaments
(see e.g. the opening exposition in \cite{chudnovsky-cycles})
it follows that $G'$ itself is acyclic.
Thus, we can find a topological ordering
$v_1, v_2, \ldots, v_n$ of $G'$.

Finally, we claim that this ordering satisfies the conditions of the
theorem.  Indeed, if an edge $(v_i,v_j)$ is positive in $G$, then
it is also an edge in $G'$, so by the property of the topological
ordering we have $i < j$.
And if an edge $(v_i,v_j)$ is negative in $G$, then $(v_j,v_i)$ is
an edge of $G'$, whence $i > j$ as required.}

\xhdr{Searching for Evidence of Global Balance and Status}
Both Theorem~\ref{thm:balance} and \ref{thm:status} have
more elaborate generalizations, when
the underlying graph is not completely connected, by generalizing
the respective three-node conditions
to arbitrary cycles.
Under these generalized conditions the conclusions remain the same:
when balance holds, we should expect to see a network that is divided
into two mutually opposed factions of friends, and
when status holds, we should expect to see a network whose
edges respect a global ordering.
We therefore take these two basic patterns ---
a division into two factions, and a global ordering --- as
potential ``signatures'' for the effects of balance and status
respectively at a global level.
Of course, at best we expect to see balance and status holding
in an approximate sense,
and so we search for approximate versions of these
two global signatures, rather than exact versions.

For balance theory, we attempt to partition the graph into
two sets to maximize the following objective function:
the number of positive edges with both ends in the same set, plus
the number of negative edges with ends in opposite sets.
(We will also say that those edges that contribute to the objective
function are {\em satisfied} by the partition.)
We develop the following maximization heuristic for this problem.
We start by randomly partitioning the nodes into two sets.
Then we repeatedly
pick a random node, and change the set it belongs to if that would
increase the value of the objective function.
We run this procedure many times from different initial starting sets.
On our datasets, we found experimentally
that the variance of the solution is very small, as the heuristic quickly
converges to a solution whose objective function value is within
a few hundred of the best solution found over all runs.
Note there is a trivial solution that would simply declare one of the sets
to be the empty set and the other to be the full node set;
this would achieve an objective function value equal to
about 80\% of the edges,
since in our datasets about 80\% of the edges are positive, and
the positive edges would be precisely those that are satisfied by
this partition.

For status theory we employ a different heuristic.
First, as in the proof of Theorem~\ref{thm:status},
we replace each negative edge with a positive edge pointing in the opposite
direction, thus obtaining
a directed network with only positive edges.
In this transformed graph, the
goal is to find an ordering of the nodes that maximizes the number of edges
pointing from 
a node earlier in the ordering to one
that is later in the ordering.
(Again, we will refer to such edges as being {\em satisfied} by the ordering.)
This is precisely the {\em Maximum Acyclic Subgraph Problem},
which is known to have strong inapproximability bounds:
a random ordering achieves an objective function value equal
to half the total number of edges in expectation, and it is computationally
hard to do asymptotically better than this in the worst case
\cite{guruswami-max-acyclic}.
Of course, our datasets are not necessarily worst-case instances ---
indeed, status theory suggests they may have additional structure ---
and we employ the following heuristic.
We start with a random ordering; we then repeatedly pick
a random pair of nodes and swap their positions in the ordering if that would
increase the value of the objective function. Again, we run this heuristic
multiple times and take the best solution found.

\xhdr{Evaluating Global Balance and Status} We now use these two heuristic
optimization methods to assess the extent to which each of the three networks
exhibits global balance and status properties.  That is, we ask whether the
quality of the partition or ordering we find is better than would be expected
from simple baselines derived from our datasets. If the quality is
significantly above such baselines, it provides evidence for these structures.
We use two such baselines. The first a {\em permuted-signs baseline} in which
we keep the structure of the network fixed, but we randomly shuffle all the
edge signs. The second is a {\em rewired-edges baseline} in which we generate a
random network where each node maintains the same number of incoming and
outgoing positive and negative edges.

\begin{table}[t]
  \small
  \begin{center}
  \begin{tabular}{l||r|r|r}
  {\bf Balance} & Epinions & Slashdot & Wikipedia \\ \hline \hline
  Netwok & 0.8344 & 0.8105 & 0.7809 \\
  Permuted & 0.8562 & 0.7779 & 0.7866 \\
  Rewired & 0.8993 & 0.8310 & 0.8316 \\  \hline
  {\bf Status} & Epinions & Slashdot & Wikipedia \\ \hline \hline
  Network & 0.7905 & 0.8221 & 0.8538 \\
  Permuted & 0.7241 & 0.7568 & 0.7767 \\
  Rewired & 0.6377 & 0.6644 & 0.6321
  \end{tabular}
  \end{center}
  \vspace{-5mm}
  \caption{Fraction of edges satisfying global balance and status.}
  \vspace{-5mm}
  \label{tab:global}
\end{table}

Table~\ref{tab:global} shows the value of the objective function (as a fraction
of the total number of edges) for each of balance and status, and across each
of our real networks in comparison to the permuted-signs and rewired-edges
baselines. For balance theory, notice that we find objective function values
that are comparable to the total fraction of positive edges, which we noted is
trivially achievable (and also trivially achievable in the two baselines, which
have the same fractions of positive edges). Moreover, if we randomize the
network structure while preserving the signed in- and out-degrees, we obtain a
network that actually achieves a {\em greater} objective function value under
our heuristics. Taken together, this suggests that there is very little
evidence for the global presence of structural balance in our three network
datasets.

For the global version of status theory, shown in the second part of
Table~\ref{tab:global}, we see quite a different picture. Roughly, we are able
to find orderings for all three datasets that satisfy about 80-85\% of all
edges. This is much higher than the 50\% obtainable from a random ordering; and
moreover, it is significantly better than the performance on either of our two
baselines. Thus, we do find evidence for an approximate global status ordering
in the real datasets, compared to baselines derived from random variations of
them.

Overall, then, there is evidence for link formation consistent with a global
status ordering on the nodes, in which positive links tend to point
left-to-right in this ordering, while negative tend to point right-to-left in
this ordering. On the other hand, we can find no significant evidence for the
kind of partitioning into factions that balance theory suggests at a global
level. This forms an intriguing contrast with our results at a local level,
where there was evidence for both balance and status. There is no contradiction
here, since the fidelity of balance and status at a local level is only
approximate, but it does raise interesting questions that suggest the need for
more powerful and general ways to relate the local structure of sign patterns
to more global forms of structure.

\vspace{-2mm}
\section{Predicting positive edges}
\label{sec:trust}
In the introduction we noted that the sign prediction problem
considered in this paper is closely related to the {\em link
prediction problem} of inferring latent relationships that are present
but not recorded by explicit links \cite{liben-nowell-link-pred}.  We
now turn to this problem in order to investigate the
role of negative edges here as well.
In particular we consider the
question of whether information about negative links can be
helpful in predicting the presence or absence of an unobserved positive link.
In other words, how useful is it to know
who a person's enemies are, if we want to predict the presence of
additional friends?

Specifically, suppose we are given a social network where the task is
to predict the presence or absence of a positive edge between two
individuals.  This is analogous to the experiments above, only now it
is the presence or absence of an edge in some context rather than the
sign of an edge that is to be predicted.  We consider two cases.  In
the first case, only information about the positive edges is used to
predict the presence or absence of a hidden positive edge, whereas in
the second case information about both the positive and negative edges
is used for the prediction.

We use the machine learning framework developed in previous sections
to build classifiers that predict whether there exists a positive edge
between a pair of nodes. We train two sets of models using the same
features (16Triads) but in one case we use the full network with
positive and negative edges, while in the other case we use
only the positive edges. We then devise the following
experiment.  For a positive edge $(a,b)$ we pick a corresponding pair
of nodes $(c,d)$ that are {\em not} connected by an edge but have the
same number of common neighbors (embeddedness) as $(a,b)$. Then we
formulate a binary classification problem where we are given a pair of
nodes and the goal is to determine whether the positive edge is
present; that is, we aim to distinguish between pairs of nodes $(a,b)$
that are connected by a positive edge and pairs of nodes $(c,d)$
that are not connected by an edge.

For each pair of nodes we compute two sets of features. For the first set of
features we include the information from positive and negative edges by
computing the features counting the frequency of each of 16 distinct signed
directed triads between a pair of nodes. For the second set of features that
are based on only positive edges we simply compute the frequencies of 4
directed paths composed of only positive edges (namely, FFpp, FBpp, BFpp,
BBpp).

We then train and evaluate two models: one trained only on features based on
positive edges, and the other trained on the whole set of features that also
include evidence coming from the negative edges. Our goal here is to understand
how information about negative relationships affects the overall performance of
predicting existence of positive edges.

\begin{table}[t]
  \small
  \begin{center}
  \begin{tabular}{l||r|r|r}
  Features & Epinions & Slashdot & Wikipedia \\ \hline \hline
  Positive edges & 0.5612 & 0.5579 & 0.6983 \\
  Positive and negative edges  & 0.5911 & 0.5953 & 0.7114
  \end{tabular}
  \vspace{-2mm}
  \caption{Predicting the presence of a positive edge.}
  \label{fig:pluspaths}
  \vspace{-8mm}
  \end{center}
\end{table}

Table~\ref{fig:pluspaths} shows the predictive accuracy for the above task.
Since we pick an equal number of edges and non-edges, random guessing has
an accuracy of 0.50. A logistic regression model using only the
features based on the positive edges improves the performance to about 0.56 for
Epinions and Slashdot, while it gives a much higher boost in Wikipedia, where
the classification accuracy almost 0.70. This is somewhat surprising as we
previously saw that for sign prediction Wikipedia was more difficult than the
other datasets.

Next we consider a logistic regression model that is trained on features based
on both positive and negative edges. This model scores 0.59 on Epinions and
Slashdot, while it improves the performance on Wikipedia to 0.71. This means
that if we use information about negative edges for predicting the
presence of positive edges in Epinions and Slashdot we get 3 percentage point
improvement in absolute terms, and a 50\% increase in the boost relative
to random guessing, compared to a
model based only on positive edges.

These results clearly demonstrate that in some settings
there is a significant improvement to be
gained by using information about negative edges, even to predict the
presence or absence of positive edges.
Thus it is often important to view positive and negative links in
an on-line system as inter-related, rather than
as distinct non-interacting features of the system.

\vspace{-2mm}
\section{Conclusion}
\label{sec:conclusion}
We have investigated some of the underlying mechanisms that determine the signs
of links in large social networks where interactions can be both positive and
negative. By casting this as a problem of sign prediction, we have identified
principles that generalize across multiple domains, and which connect to
social-psychology theories of balance and status. Moreover, our methods for
sign prediction yield performance that significantly improves on previous
approaches. At a global level, we have seen that there is evidence in all of
our datasets for an approximate global status ordering on nodes, while we find
in contrast that there is essentially no evidence for a global organization of
these networks into opposing factions; this suggests that balance is operating
more strongly at a local level than at a global one. Finally, we have seen that
employing information about negative relationships can be useful even for tasks
that involve only the positive relationships in the network, such as the
problem of link prediction for positive edges.

There are a number of further directions suggested by this work. A first one is
of course to explore methods that might yield still better performance for the
basic sign prediction problem, and to understand whether the features that are
relevant to more accurate methods help in the further development of social
theories of signed links. We are also interested in strengthening the
connections between local structure and global structure for signed links.
Finally, as noted at the outset, the role of positive and negative
relationships in on-line settings is not limited to domains where they are
explicitly tagged as such.

\xhdr{Acknowledgements} We thank Michael Macy for valuable discussions, and the
anonymous reviewers for a number of comments and suggestions that helped to
improve the paper. Research was supported in part by the NSF grant IIS-0705774, IBM
Faculty Award, gift from Microsoft Research and Yahoo! Research Alliance grant.

\bibliographystyle{abbrv}


\end{document}